# Hidden in-plane long-range order in an amorphized crystal

Yin Chen,[1]† Anthony E. Phillips, [2]† Cheng Fu,[3]† Volodymyr Bon,[4]† Lei Liu,[5,6]† Xiaoxu Sun,[1] Jiahui Wang,[1] Na Lin,[1] Ruize Xie,[1] Guanqun Cai,[7] Yutong Wang,[4] Jing Ma,[3] Yuhong Liu,[5] Yu Han,[8] Stefan Kaskel[4]

[1] Central South University, College Chemistry & Chemical Engineering, Changsha 410083, Hunan, China
[2] Queen Mary University of London, School of Physical and Chemical Sciences, London E1 4NS, London, UK
[3] Nanjing University, College of Chemistry & Chemical Engineering, State Key Laboratory of Coordination Chemistry, Nanjing 210008, Jiangsu, China
[4] Technische Universität Dresden, Department of Inorganic Chemistry I, D-01069 Dresden, Germany
[5] Tsinghua University, State Key Laboratory of Tribology in Advanced Equipment, Beijing 100084, Beijing, China
[6] School of Mechanical Engineering, North China University of Water Resources and Electric Power, Zhengzhou 450045, PR China
[7] Sichuan University, Institute of Atomic & Molecular Physics, Chengdu 610065, Sichuan, China
[8] South China University of Technology, Center for Electron Microscopy, Guangzhou 510640, Guangdong, China

† These authors contribute equally.

**Abstract**

Solid materials are commonly classified as crystalline or amorphous based on the presence or absence of long-range order. Metal-organic frameworks (MOFs), like other solids, also display markedly different properties and functions in these two phases. Here, we identify a previously unrecognized structural state that retains long-range in-plane translational order while losing order along the stacking direction. Hypothesized since 1941 but not experimentally verified, this intermediate phase emerges in a crystalline MOFs via controlled thermal desolvation, which selectively disrupts the intrinsically weak interlayer interactions while preserving macroscopic structural coherence. Although the resulting material appears amorphous under conventional characterization, systematic synchrotron PXRD, total X-ray scattering, and low-dose high resolution-TEM reveal clear in-plane periodicity. This material spontaneously delaminates in water into uniform, high-quality two-dimensional crystalline nanosheets, forming stable colloidal suspensions and exhibiting superlubricity comparable to graphene – but at less than 0.1% of the production cost. Our discovery finds a missing link within the long-standing crystalline-amorphous dichotomy, while providing an inherently scalable route to high-quality 2D crystals, and offering a conceptual and practical advance in phase engineering.

**Main:**

Phase engineering enables the design of materials with tailored properties and functions[1-3]. Disordered or partially crystalline materials can demonstrate functionalities that differ dramatically from both near-perfect crystals and amorphous materials of the same composition[4-7]. When dimensionality is reduced from bulk to two-dimensional (2D) [8], the resulting 2D crystals exhibit exceptional mechanical, electronic, optical, catalytic, separation and biological properties[9-13]. However, a long-standing challenge persists: scalable production of 2D crystals without compromising in-plane crystallinity. In principle, only weak van der Waals and other interlayer forces must be overcome. In practice, dense interaction sites act cooperatively, requiring harsh chemical or physical treatments. Liquid-phase exfoliation, the most widely used large-scale fabrication method[14-16], typically relies on surfactants, highly polar solvents, or intercalants to weaken or counteract interlayer interactions[17,18]. Yet supramolecular preorganization within 2D crystals is thermodynamically far more stable than that of interactions with small molecules[19]. As a result, 2D materials obtained by conventional approaches often suffer from poor structural quality, high defect densities, limited lateral-sizes, severe aggregation and partial recrystallization, with practical performance far below theoretical expectation.

Here, we report a fundamentally different approach that leads to the discovery of a previously unidentified phase. By drastically reducing the intrinsic interlayer cohesion of layered solids, thermal fluctuations alone become sufficient to induce spontaneous interlayer slippage, leading to out-of-plane disorder, or amorphization along single dimension. The resulting state is a hybrid of long-range order and disorder, fundamentally distinct from conventional "stacking faults", which are commonly observed in different layered materials such as graphite[20], molybdenum disulfide[21], perovskites[22], covalent organic frameworks (COFs)[23,24], ice I[25] and even diamond[26]. In these materials, the in-plane translation between consecutive layers is randomly selected from a limited set of possibilities, constituting a form of correlated disorder (Extended Data Fig.1)[27]. In contrast, thermodynamic disorder along the stacking direction corresponds to the "random layer lattice" model proposed by B. E. Warren, an intermediate state long-hypothesized to bridge crystalline and amorphous phases[28,29]. Conceptually, it emerged alongside liquid crystals[30] and plastic crystals[31], and decades before quasicrystals[32], yet has not been experimentally realized in a well-defined bulk material[33]. Previous few experimental hints are restricted to nanoparticulate COFs [24,34], which also exhibit poor intralayer crystallinity[35], or to nanometer-thick CVD-grown films[36,37], both of which differ substantially from this theoretically envisioned intermediate phase.

MOFs offer highly tunable composition and architecture[38,39], providing an ideal platform for engineering interlayer interactions and accessing structure with tailored disorder [40,41]. Guided by molecular design, we constructed a layered MOFs, **IPM-1** (Intermediate Phase Material), from cage-like building block molecules. **IPM-1** features an ultralow areal density of interlayer bonding sites (~ 0.2 site/nm$^2$). Structural modeling and first-principles calculations found that its interlayer cohesion is dominated by solvent molecules confined within the lattice channels; upon their removal, the remaining interactions become sufficiently weak that thermal fluctuations alone can overcome them. The resulting material falls outside conventional structural classifications: although it appears X-ray amorphous under routine characterization, systematic synchrotron and electron microscopy analyses reveal that its intrinsic in-plane periodicity is preserved. The apparent amorphization arises from the extremely weak 2D diffraction signals, which lie below the detection limits of standard laboratory techniques. Consistent with theoretical expectations, **IPM-1** spontaneously delaminates in water to yield uniform, high-quality 2D nanosheets—without any chemical additives or external driving forces. Overall, this work overcomes long-standing experimental barriers to realizing the long-hypothesized intermediate phase and enables facile access to high-quality 2D materials. The discovery of this previously unrecognized structural state in MOFs provides new conceptual and practical insights into the design of framework materials and other layered materials.

**Results and discussion**

**Highly "amorphizable" yet stable IPM-1 crystal**

**IPM-1** was synthesized by reacting bicyclocalix[2]arene[2]triazine tricarboxylic acid (BCTA) with $MnCl_2$ in DMF (N,N-dimethylformamide) on a gram-scale with a yield of ~80%[42]. The product forms transparent crystals with well-defined morphologies (Fig. S1). Single-crystal analysis reveals a layered and porous structure, formulated as $[BCTA]_2Mn_3(DMF)(H_2O)_3$. The compound crystallizes in the triclinic space group $P\bar{1}$, with unit cell parameters $a$=21.42 Å, $b$=22.62 Å, $c$=23.23 Å, $\alpha$=61.17°, $\beta$=64.62°, $\gamma$=74.95°. Each layer adopts a wave-like honeycomb arrangement (Fig. 1A, Data S1) stacked in an ABAB sequence (Fig. 1B and S2). Two distinct Mn clusters are present: $[Mn_3(O_2C)_6]\cdot 4H_2O$ and $[Mn_3(O_2C)_6]\cdot 2H_2O\cdot 2DMF$ (Fig. S3). Interlayer cohesion arises from weak O–H···O hydrogen bonding between these clusters, with only one such site per ~5 nm² (Fig. S4). As a result, the effective contact area and van der Waals interactions between adjacent layers are far lower than in graphite.

Surprisingly, in-house powder X-ray diffraction (PXRD) consistently revealed that as-prepared **IPM-1** crystals exhibit an amorphous pattern, indicating an unusually high tendency for amorphization. Freshly isolated crystals—without any washing or drying treatment— showed diffraction patterns consistent with the single-crystal structure (Fig. 1C). Upon heating under a nitrogen flow (2 mL/min), Bragg peaks began to weaken at 60 °C and completely disappear after 2 h, indicating a transition to an amorphous state (Fig. 1D). At 120 °C, this transformation occurred within only 15 min (Fig. S5). Optical microscopy revealed no apparent morphological changes even after 30 min at 140 °C (Fig. 1E). These results demonstrate that the loss of long-range order occurs without structural collapse, especially under such mild conditions. While desolvation-induced amorphization has been reported before, it typically involves structure degradation[43]; the structural nature of such rare exceptions remains unclear[35].

Further characterizations confirmed its remarkable structural stability. Thermogravimetric (TG) analysis showed a weight loss of ~11 wt% below 150 °C (Fig. S6), attributed to the release of DMF molecules confined within the channels. Significant thermal decomposition was only observed at temperatures exceeding 400 °C. TG–MS (Mass Spectrometry) and TG–FTIR (Fourier Transform Infrared Spectroscopy) analyses confirmed solvent release starting above 50 °C and peaking near 100 °C (Fig. S7–S8), consistent with the PXRD results. After vacuum activation at 120 °C, Brunauer–Emmett–Teller (BET) analysis revealed a surface area of 210 m²/g and a pore size distribution matching the single-crystal structure. High-pressure sorption experiments demonstrated fully reversible $N_2$ adsorption–desorption cycles even under 200 atm (Fig. S9–S10). **IPM-1** also exhibits a $CO_2$ uptake capacity of 1.3 mmol/g, while in-situ PXRD confirmed that its structure remains unchanged during the cycles (Fig. S11) [44]. Importantly, partial recovery of original crystallinity was achieved: after 4 h desolvation under 80 °C, amorphized **IPM-1** crystals partially regained crystallinity upon treatment in DMF at 120 °C for 48 h, as evidenced by the reappearance of low-angle PXRD reflections (Fig. S12).

**The hidden long-range in-plane periodicity**

We hypothesize that this anomalous amorphization arises from extremely weak interlayer cohesion, leading to random disorder between layers. In theory, diffraction signals from such a random layer lattice could be too weak to detect; however, direct experimental evidence has been lacking[28]. To clarify this behavior, the desolvation process was monitored *in situ* using synchrotron radiation (SR) PXRD at the BL14B1 beamline of SSRF ($\lambda$ = 0.6887 Å). Freshly separated **IPM-1** samples, sealed in capillaries on-site, initially exhibited a series of sharp, intense peaks (Fig. 2A), fully consistent with the simulated pattern, confirming the high crystallinity of **IPM-1** in mother liquor. Upon heating to 80 °C for 5 min, the SR-PXRD pattern changed markedly, indicating a phase transition: the original *hkl* reflections weakened substantially and disappeared at higher scattering angles, whereas a new set of weak peaks emerged. At temperatures above 100 °C, these *hkl* reflections completely vanished, leaving only four peaks corresponding to a quasi-2D phase. Notably, no further structural changes were observed upon cooling (Fig. S13).

Pawley refinement was performed with the newly observed diffraction peaks using a set of transformed lattice parameters (Fig. 2B), in which the 2D layers lie within the *ab*-plane and are stacked along the *c*-axis, allowing a clearer and more intuitive interpretation of the structural evolution (Data S2). The lattice retains the same space group P$\bar{1}$, but the unit cell parameters were refined to *a*=42.92 Å, *b*=23.33 Å, *c*=21.41 Å, *α*=100.07°, *β*=30.60°, *γ*=115.63°. The weak peaks at 1.90°, 2.01°, 2.19° were assigned to the 1$\bar{1}$, 01 and 10 *hk*-type reflections respectively. The broad and strongest peak observed at 3.48° (corresponding to d-spacing of 1.13nm) can be assigned as the 001 reflection, consistent with the thickness of a single layer (Fig. 1A). These *hk*-type reflections were still observed in the SR-PXRD pattern of **IPM-1** nanosheets, albeit with very poor intensity, likely due to the reduced size and increased flexibility of the free-standing nanosheets. However, the 001 reflection completely disappeared (Extended Data Fig.2), which confirms that turbostratic disorder rather than interlayer delamination is formed during the desolvation of **IPM-1**. These observations indicate that the phase transition arises from the dimensionality reduction of the origin three-dimensional lattice.

Variable-temperature X-ray total scattering experiments were further conducted at the I15-1 beamline at DLS (λ= 0.161669Å), and pair distribution functions (PDF) were extracted [24,34]. As expected, the sharpest two peaks occur at low *r*: the first at 1.36 Å corresponding to C-C, C-N and C-O bonds within the organic ligands; the second at 1.8-2.4 Å corresponding to the Mn-O coordination bonds (Fig. 2C, S14). The pattern at higher *r* is complex to fully model due to the very large unit cell, but can be qualitatively attributed primarily to Mn...Mn interactions, given that Mn possesses the largest scattering length among the constituent elements. The experimental PDF results support this point, showing good agreement with the calculated Mn…Mn peaks at high *r* region (Fig. S15), confirming the existence of long-range order in **IPM-1** lattice after desolvation. Each layer contains "triplets" of Mn atoms (Fig. 2D, Extended Data Fig.3). If random interlayer disorder occurs at elevated temperatures, peaks corresponding to *inter*-layer correlations–such as those labelled B, D and E–would be expected to broaden substantially, while *intra*-layer correlation peaks, such as A and C, would remain sharp. Consistent with this expectation, characteristic broadening of peaks B, D and E is observed, particularly at the two highest temperature points, while peaks A and C remain essentially unchanged, but shift slightly to a lower *r* value with increasing temperature. These results align with SR-PXRD observations and confirm that **IPM-1** loses its order only along the stacking direction.

**Mechanism of the dimensionality reduction phase transition**

Now, we have established that amorphization observed in **IPM-1** is fundamentally distinct from conventional cases, occurring exclusively along one dimension. To understand this process, density functional theory (DFT) calculations were performed to quantify the interlayer interactions in **IPM-1** and elucidate the mechanism of this transition.[45] The ground-state energy difference between stacked layers and an isolated monolayer was computed (Fig. 3A, Table S2, Data S3). The DFT optimized crystal structure exhibited a better fit to the experimental PXRD pattern at high angles compared to the single-crystal structure, further confirming the high crystallinity of **IPM-1** (Fig. S16). Unexpectedly, the interlayer binding energy was determined to be 2.18 meV/Å$^2$ with the single crystal structure (Table S3), suggesting that aggregation of layers is energetically *unfavorable* compared to delamination. Experimental results also show a slightly negative energy as -1.10 meV/Å$^2$ for exfoliating a single **IPM-1** layer from the bulky crystal (Table S4, Fig. S17–19), further confirming a spontaneous delamination. For comparison, the exfoliation energy of graphene was calculated to be 18.36 meV/Å$^2$ using the same method[46]. These counterintuitive results explain why interlayer disorder occurs readily but do not clarify how **IPM-1** can still be prepared in high yield.

Notably, the above calculations exclude DMF molecules disordered within the lattice, omitted during single-crystal structure refinement. Elemental analyses indicate approximately seven disordered DMF molecules per Mn cluster (Table S1). To understand their effect, binding energies were calculated according to the formula $E_{bind} = E_{Total} - E_{layer} - E_{(DMF)n}$ for two models with differing numbers of solvent molecules (2 or 7 DMF per Mn cluster). The binding energies were found to be -2.03 meV/Å$^2$ and -23.19 meV/Å$^2$ (Table S5, Fig. 3B), respectively, indicating that freshly prepared, solvated **IPM-1** crystals have stronger interlayer binding energy than graphite (-21 meV/Å$^2$). This finding underscores the critical role of DMF solvent in stabilizing the 3D-lattice through enhanced interlayer interactions, which even become negative after the removal of DMF molecules.

Molecular dynamics (MD) simulation (see SI, Data S4) further confirmed that slippage between desolvated **IPM-1** layers is thermodynamically spontaneous. Adjacent layers oscillate around the equilibrium position, with an energy variation of approximately 1.2 meV/Å$^2$ and a relative displacement up to 2 Å (Fig. 3C, 3D). This displacement is large enough to render all translations within the *ab* plane between adjacent layers equally probable. The maximum energy barrier observed on the potential energy surface is less than 7.0 meV/Å², notably lower than that of graphene. The coordinated DMF of the Mn cluster plays a key role in stabilizing the lattice. Without it, the calculated relative displacement can reach 2-3 nm, which is sufficient to induce physical collapse of the crystal (Fig. S20–24).

Taken together, these DFT calculation results reveal that this unprecedented amorphization originates from the extremely weak interlayer cohesion imposed by the cage-like BCTA building blocks and large porous channels. Solvent molecules bridging adjacent layers provide essential stabilization, leading to its high-yield preparation; once removed, interlayer interactions collapse, enabling spontaneous sliding and amorphization along a single dimension.

**Self-exfoliation towards high-quality nanosheets**

With the interlayer interactions now overcome by thermal motion, the next question is whether large-scale preparation of high-quality 2D crystals can be achieved. Remarkably, **IPM-1** crystals spontaneously fully dispersed in water within one minute, whereas the free BCTA ligand is entirely insoluble in water (Movie S1). The suspension appeared light milky white and displayed a clear Tyndall effect (Fig. 4A). As shown by the scanning electron microscopy (SEM) images (Fig. 4B and S25), the solution contains a large number of ultrathin nanosheets with exceptionally smooth surfaces, indicating that **IPM-1** undergoes self-exfoliation rather than decomposition, forming a stable and uniform aqueous nanosheet suspension. The exfoliation behavior proved solvent dependent. **IPM-1** crystals do not dissolve in methanol, but still exfoliated within a few minutes (Movie S2, Fig. S26). Ethanol induced slower exfoliation (Movie S3, Fig. S27), while in dichloromethane, no morphological change was observed even after 24 h. PXRD confirmed amorphization similar to that induced by heating (Fig. 1C and S28). These results show that strong protic solvents disrupt the weak interlayer interactions in **IPM-1**, enabling thermal motion of solvent molecules to drive spontaneous exfoliation into uniform 2D nanosheets.

Nanosheets derived from **IPM-1** preserved perfect in-plane crystallinity. High-resolution transmission electron microscopy (HR-TEM) and atomic force microscopy (AFM) images revealed that these nanosheets have lateral sizes up to tens of micrometres and a uniform thickness (Fig. 4C, 4D, S29–31). Low-dose HR-TEM clearly confirms that the nanosheets retain the same 2D periodicity as the original crystal [47]. The three interplanar spacing values observed in HR-TEM image show excellent agreement with the d-spacing values detected by SR-PXRD as 2.06 nm ($1\bar{1}$), 1.99 nm (01) and 1.83 nm (10), respectively (Fig. 2G, S32–33). IR spectra of the nanosheets remained identical to those of the pristine crystal, while BET analysis revealed a nearly doubled surface area (Fig. S34–35) and sharp pore-size distributions at approximately 0.7 nm, 1.3 nm, and 1.8 nm, respectively (Fig. S36), confirming its exceptional structural stability and in-plane crystallinity.

One key application of 2D materials is their use as solid superlubricants (friction coefficient < 0.01), offering promising potential in microelectromechanical systems [48]. Achieving this performance requires large, defect-free 2D crystal and incommensurate interlayer contact [49]. Graphene and $MoS_2$ are the most prominent solid superlubricants to date. However, only high-quality nanosheets prepared by scotch-tape exfoliation can achieve this, obtained exclusively from highly oriented pyrolytic graphite (HOPG) or from large-size single crystals of $MoS_2$. Recent studies reported that 2D MOFs, such as CuTCPP, also exhibit superlubricity. But this kind of high-quality 2D MOFs can only be prepared by confinement methods, with yields typically below 1 μg per batch and substantial thickness[50]. We have employed our previously developed method to evaluate the superlubricant performance of self-exfoliated **IPM-1** nanosheets. Both a $SiO_2$ microsphere probe slider and a monocrystalline silicon substrate were coated with the nanosheet suspension to form a homojunction (Fig. S37–38). The nanosheets tiled uniformly with low roughness across the contact area (Fig. S39). Initially, the homojunction exhibited relative sliding over a 3 μm × 3 μm area under varying loads. The frictional force increased linearly with the applied load (Fig. 4F), yielding a friction coefficient of 0.0057, indicative of a superlubricity state comparable to that of high-quality graphene or $MoS_2$ nanosheets. The interfacial adhesion of the homojunction, specifically the interlayer force, was calculated as 19.4 nN, corresponding to 57.6% of that of graphite (Fig. 4G, S40)[51]. The **IPM-1** homojunction maintained a stable superlubricity state for over 30 minutes without any detectable wear on the substrate surface (Fig. S41). These results confirm the high quality and performance of the self-exfoliated **IPM-1** nanosheet, achieved at a production cost less than 0.1% of that of scalably produced graphene (Table S2)[52].

**Discussion**

These experimental results fully validate our hypothesis: the intrinsic interlayer cohesion of layered solids can be engineered through molecular design, enabling a thermodynamically driven and scalable pathway to high-quality 2D crystals. Beyond its synthetic implications, this finding carries profound theoretical significance. **IPM-1** cannot be classified as either conventional a crystalline or an amorphous solid, but rather as an intermediate structural regime that bridges these two fundamental states. It provides direct insights into amorphization mechanism and the complex structural order arising in disordered materials[4,5,41]. Importantly, previous studies have shown that many structural motifs self-assemble into layered frameworks with inherently weak interlayer interactions[53-57], suggesting that a broader family of intermediate-phase materials may exist beyond **IPM-1**. Furthermore, X-ray amorphous products frequently emerge during the synthesis or post-treatment of frameworks and porous materials[10,58,59], while desolvation-induced amorphization is commonly observed[35,60,61]. The melting and glass formation of MOFs has even become a particularly active research topic[6,62,63]. Yet, the underlying structural evolution and mechanisms of these transformations remain largely unresolved. Our findings suggest that many materials conventionally dismissed as "amorphous" may in fact possess hidden long-range order, rendering them unrecognized precursors for high-quality 2D crystals. This work therefore calls for a reassessment of the amorphous state–not as a terminal disorder, but as a latent structural phase space with untapped potential for functional materials design and phase engineering.

**Acknowledgments:** We acknowledge the provision of beamtime by Diamond Light Source (DLS), Shanghai Synchrotron Radiation Facility (SSRF) and Deutsches Elektronen Synchrotron (DESY). YC thanks Prof. Mei-Xiang Wang, Martin T. Dove, Andrew L. Goodwin, De-Xian Wang, Haibo Zhu, David A. Keen, Yong Cui, Minghua Liu, Xiaoyi Yi, Haiyu Hu, Kostya Trachenko, Xiaolong Li, He Lin, Tianping Ying, Qikui Liu for helpful discussion and support, and the School of Physical and Chemical Sciences at Queen Mary University of London for hosting a research visit.

**Funding:** This work was supported by State Grid Shaanxi Electric Power Research Institute (to YC), EPSRC (EP/S03577X/1 to AEP), NSFC (Grant 22033004 to JM, 52350323, 52488101 to YHL), NSFC Excellence Research Group (Grant 52488101 to YHL) Deutsche Forschungsgemeinschaft (DFG CRC 1415 to SK) and China Postdoctoral Science Foundation (to LL).